\documentclass[aps,PRL,twocolumn,superscriptaddress]{revtex4-2}
\usepackage{graphicx}

\usepackage{color}
\usepackage{latexsym}

\usepackage{xcolor}
\usepackage{hyperref}
\usepackage{amsmath,amssymb}
\usepackage{float}
\usepackage{bm}
\usepackage{tikz}
\usetikzlibrary{shapes}

\begin{document}
\title{Anomalous relocation of topological states }
\author{Hamidreza Ramezani}
\email{hamidreza.ramezani@utrgv.edu}
\affiliation{Department of Physics and Astronomy, University of Texas Rio Grande Valley, Edinburg, TX 78539, USA}



	\begin{abstract}
I show that a single embedded non-Hermitian defect in a one-dimensional topological system at certain degrees of non-Hermiticity can remove the topological mode from the edge and restore it inside the lattice at the same place where the non-Hermitian defect is placed. I relate this unexpected phenomenon to the wave matching condition and continuity of the wave function at different sites in the lattice. These findings pave the way for controlling the position of topological states at will.
\end{abstract}

\maketitle

\emph{Introduction}--- In recent years topological systems became the center of attention due to their peculiar robustness against disorders. Due to this interesting feature topological systems found their ground in different areas of physics, and engineering, including condensed matter physics \cite{Has10,Qi11}, photonics \cite{Lu14,Oza19}, Floquet systems and quantum walks \cite{Kit12,Rec13b}, ultracold atomic gases \cite{Gol16}, acoustics \cite{Yan15,Peng_2016, R-F2022}, mechanics and robotics \cite{Kan14,Hub16,Gha19}, and electronics \cite{Lee18}.

The robustness of topological systems is restricted in a state that is located at the edge of a 1D and 2D topological lattice or the surface of a 3D topological lattice. Recently, the study of topological systems has extended to non-Hermitian systems where the interplay between non-Hermiticity and topology results in unexpected phenomena \cite{Lu14,Khanikaev_2017,Longhi_2017,Peng_201611,Sch13,Mal15,Pol15,Zeu15,Zhao_2015,El_Ganainy_2015,Leykam_2017,Wei17,Yuce_2022}. For example, non-Hermiticity has been incorporated to induce a pure imaginary zero mode in the middle of a structure \cite{Pan18,F_mos20}. The combination of the non-Hermitian skin effect, the pile-up of bulk states at the edge due to the non-Hermiticity, and the topological edge state can lead to the delocalization of the topological edge state \cite{Longhi:2018aa,Yuce_2020, Zhu_2021}.

As is shown in the aforementioned and related works, it is believed that a topological state is always intact and remains robust in its original position as long as the topology of the system remains the same. Thus, inserting a single defect far away from the edge will not affect the position of a topological state as such a single hermitian or non-Hermitian defect cannot change the topology. This picture seems to be compatible with the fact that topological states are exponentially localized and thus, a single defect far away from the edge has minimal effect on the localized state sitting at the edge. Here I challenge this common believe and show that while a single defect cannot change the topology it can remove a topological state from its original position, namely the edge, and place it somewhere inside the lattice. 
\begin{figure}
	\centering
	\includegraphics[width=\columnwidth]{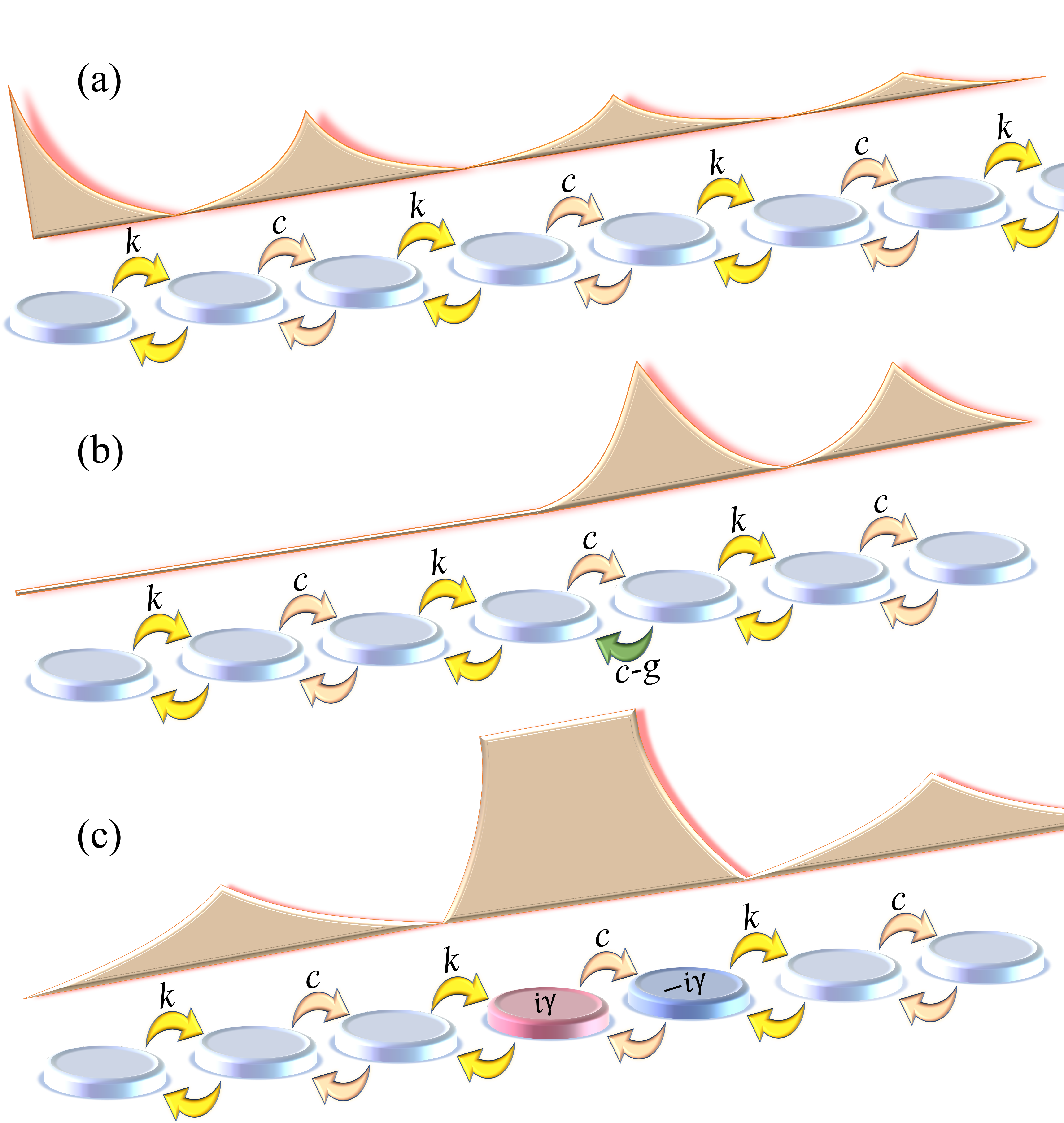}
	\caption{Schematic of topological state relocation in coupled resonator lattices. Lattice with (a) no defect, (b) with a non-Hermitian coupling defect, and (c) with a PT-symmetric defect. The couplings are selected as $k<c$ resembling a topological SSH chain. In (a) we expect a topological localized state on the left edge. In (b) one of the intra-dimer couplings is not symmetric namely in one way it is $c$ and the other way is $c-g$. For a range of values of $g$ the topological state disappears from the edge and re-appears in one of the sites next to the defective coupling. In (c) the topological state again disappears from the edge and re-emerge in the PT-symmetric defect. for a range of values of gain and loss parameter $\gamma$. }
	\label{fig1}
\end{figure}

In this Letter, I show that without changing the topology of a system and by incorporating {\bf only one single non-Hermitian defect} in a Hermitian or non-Hermitian topological lattice one can relocate a topological state from the edge and place it inside the lattice. More specifically, in a Su-Schrieffer-Heeger (SSH) lattice I show that for small values of non-Hermiticity in the defect the topological state stays at the edge. However, by increasing the non-Hermiticity in the defect the intensity of the topological state, in its spatial representation, reduces at the edge and increases at the defect state. This process continues until the strength of the non-Hermiticity becomes equal to the strength of the larger coupling in the lattice where the topological state finds its maximum intensity around the defect. At this point, the topological state is no longer at the edge. By increasing the non-Hermiticity strength to go beyond the coupling the reverse process starts where the topological state goes back to its original form, namely having its maximum intensity at the edge. 
I show that the wave matching condition can reveal the physics behind the relocation of the topological state. Notice that there is no correlation between the position of the defect and the topological edge state. Thus, we can put the defect much further away from the edge and still observe the relocation of the topological state. The only requirement to have the relocation is to introduce the defect in a dimer with a larger coupling and increase its strength to reach the larger coupling. This relocation of the topological state is robust against disorder and always occurs for some values of the non-Hermiticity in the defect state.  
This proposal opens a new direction for manipulating the topological states.

\emph{Model}--- While my proposal can be incorporated in a variety of different systems including photonics  \cite{StJ17,Zha18,Par18,Ota18, Rec13b,Zeu15}, acoustics \cite{Yan15,Peng_2016,Zan19, Puri_2021, R-F2022}, and electronics \cite{Lee18}, here, I consider a one-dimensional lattice composed of $ {\cal N}\in odd$ sites. As shown schematically in Fig.(\ref{fig1}), I assume that the lattice sites are evanescently coupled microdisk resonators \cite{Cao15} arranged such that it calls for an SSH model \cite{Su79}. For simplicity, I consider the case where the real part of the resonance frequency of the coupled modes is $\omega_0$, while the imaginary part is zero. Without loss of generality, I set $\omega_0=0$.
The eigenvalue problem associated with the coupled-mode equation that describes this system in the $n^{th}$ microdisk is given by
\begin{equation}
\begin{array}{cc}
E\psi_{n}=   k \psi_{n+1},\quad n=1& (a)\\
E \psi_{n}= k \psi_{n-1} + c \psi_{n+1}, \quad n\in even &(b) \\
E \psi_{n}= c \psi_{n-1} + k \psi_{n+1}, \quad n(\neq {\cal N})\in odd&(c)\\
E \psi_{n}=   c \psi_{n-1},\quad n={\cal N}&(d)
\end{array}
\label{eq1}
\end{equation}
where $\psi_{n}$ is the modal field amplitudes in a disk, $k$ and $c$ are couplings between the adjacent disks that are real and fulfill $k<c$. The choice of $0<k<c$ in Eq.(\ref{eq1}) makes sure that hybridization of the sites occurs between adjacent disks connected by coupling $c$, I refer to them as a dimer. Thus there exists a non-trivial topological state at the left edge of the lattice with eigenfrequency $ E=E_T=0$ as represented schematically in Fig.(\ref{fig1}a).

To observe the topological mode relocation here I consider two different types of defects each capable of repositioning the topological mode. In one case I will replace one of the $c$ couplings with an asymmetric coupling while in the second one I add a balance gain and loss to one of the dimers in the lattice. The relocated topological mode will have different properties in each case.
\begin{figure}
	\centering
	\includegraphics[width=1.\columnwidth]{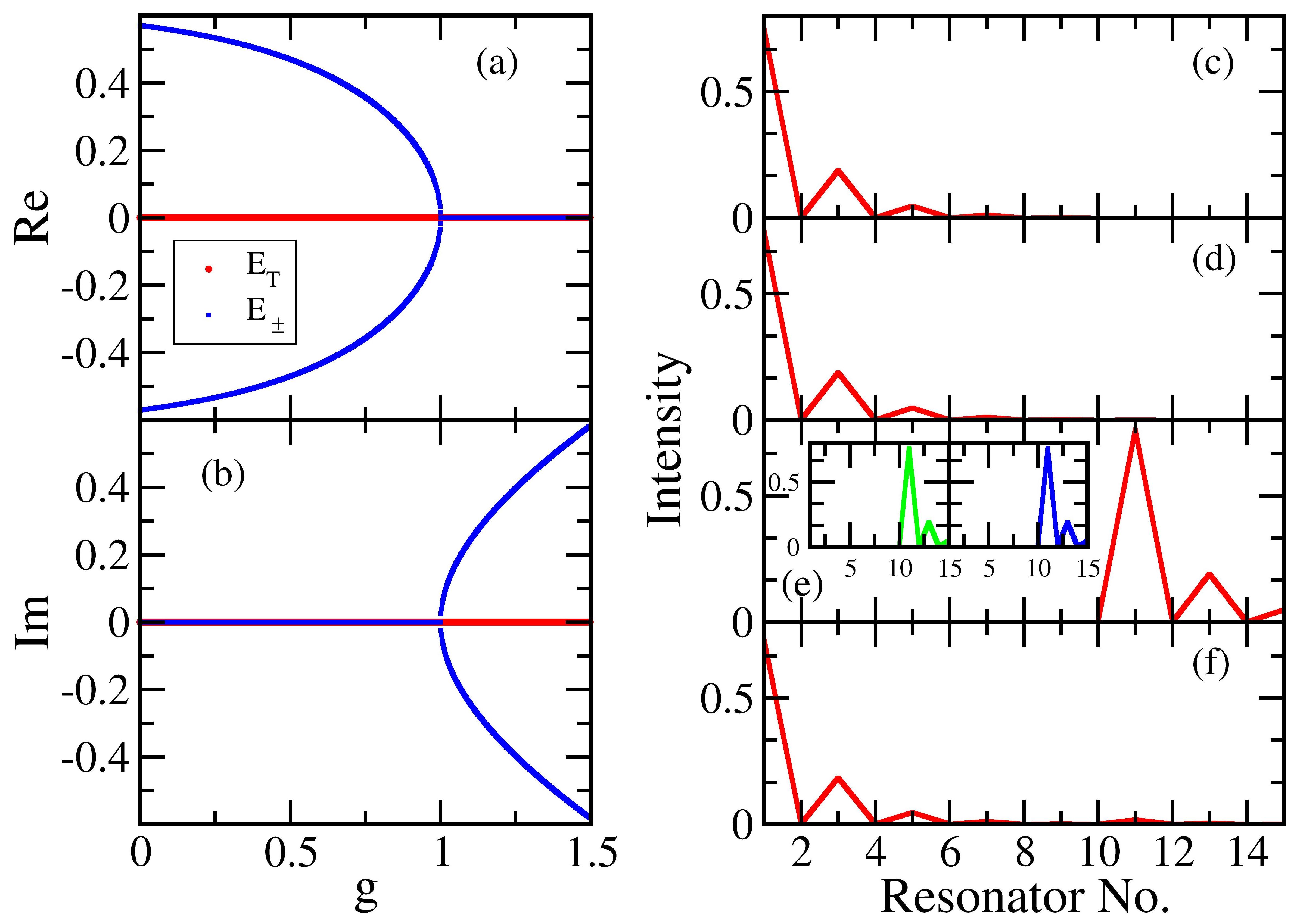}
	\caption{Numerical demonstration of the topological mode relocation via an asymmetric coupling defect shown in Fig.(\ref{fig1}b). (a) Real part of the energy associated with the topological state, $E_T$ (red dots) and the bound states $E_{\pm}$ (blue dots) as a function of $g$ parameter. (b) The same as (a) but the imaginary parts of the energies. We normalized all the couplings to $c=1$ and chose $k=0.5$. As $g$ approaches $c$ lattice gets close to a second-order exceptional point. (c,d) The intensity of the topological state versus resonator numbers when $g=0$ and $g=0.5$, respectively. In both cases, the topological state is located at the edge. (e) For $g=c$ the topological state is relocated to the site $\psi(2m+1)$ with no intensity at prior sites. The insets show the intensity of the bound states with $E=E_{\pm}$. (f) For $g=1.5>c$ the topological state goes back to the left edge.}
	\label{fig2}
\end{figure}

\emph{ Asymmetric coupling}--- Let us now change one of the Hermitian couplings $c$ in the lattice depicted in Fig.(\ref{fig1}a) such that state $\psi_{2m}$ and $\psi_{2m+1}$, corresponding to a dimer in the lattice, become asymmetrically coupled to each other, resulting in a lattice shown in Fig.(\ref{fig1}b). Here, $m$ is an integer number that its value can be selected once from the range $1\le m\le \frac{{\cal N}-1}{2}$. Thus, from now on, in this notation, $m$ depicts the position of the defect in the lattice. In this case, while the equations describing the eigenfrequencies in all other microdisks remain the same as the ones given by Eq.(\ref{eq1}), the eigenvalue equations associated with the disks coupled through the non-Hermitian coupling will be given by
\begin{equation}
\begin{array}{c}
E \psi_{2m}= k \psi_{2m-1} + (c-g) \psi_{2m+1}\\
E \psi_{2m+1}= c \psi_{2m} + k \psi_{2m+2}\\
\end{array}
\label{eq2}
\end{equation}
In the above equation, $g$, which we assume is larger than zero, describes the amount of asymmetry in the coupling. While for $g=0$ this structure goes back to the Hermitian one shown in Fig.(\ref{fig1}a), any deviation of $g$ from zero results in the jumping of two modes from bands to the gap, where each mode corresponds to a non-topological localized state.  As shown in the Fig. (\ref{fig2}a,b), the non-topological localized` modes have real eigenfrequencies $E=E_\pm$ in the exact phase which for a reasonably large lattice occurs for $g<c$ and becomes complex for $g>c$ with eigenfrequencies $Re(E_{\pm})=0$ and $Im(E_{\pm})\neq0$. At $g\approxeq c$ we will have a second-order exceptional point in which two eigenvalues and their corresponding eigenvectors of the Hamiltonian associated with the system become degenerate. For $g>c$ these $E_{\pm}$ modes have robust features and their real part remains zero as long as they are in the broken phase \cite{F_mos20}. On the other hand, the original topological mode $E_T$ remains real irrespective of the value of $g$ as shown in Fig. (\ref{fig2}a,b). Thus, one can claim that the single defect $g$ does not change the topology of the lattice. 

To see the effect of the single defect $g$ on the topological mode, in Fig. (\ref{fig2}c-f) I have plotted the mode profile associated with the original topological mode with $E_T=0$ for various $g$ values. Specifically, at $g=0$ and $g=0.5$ in Fig. (\ref{fig2}c,d) respectively, we clearly have the expected topological mode that appears on the left side of the lattice. Note that I intentionally put the defect far away from the left edge of the lattice, namely in between the site number $10$ and $11$, where the topological mode has almost zero intensity. In figure (\ref{fig2}e) I chose $g=c$. Surprisingly, we observe that the topological mode is no longer localized at the left edge and instead appears in the same place where the defect coupling is located. Note that as depicted in the insets of the same figure the defect modes with energy $E_{\pm}$ are appearing in the same place. By increasing the $g$ to larger values, as seen in Fig.(\ref{fig2}f) for $g=1.5$, the topological mode restores its original position at the left side of the lattice.

The disappearance of the topological mode from the edge is an unexpected phenomenon. If the value of $g$ could change the topology firstly we should see that its zero energy eigenvalue $E_T$ disappears, which is not the case depicted in Fig(\ref{fig2}a), and at the same time, no localized mode should exist. The topological mode indeed disappears from the edge however, it reappears at the same place that the defect exists. Furthermore, this re-localization is not significant for all values of $g$ and it becomes apparent for a small range of $g$-values around the $g=c$. Note that the exceptional point for the two localized defect modes associated with $E_\pm$ occurs at $g\approx c$.

At this point let me explain the nontrivial relocation of the topological mode at $g=c$ for a semi-infinite lattice using the wave matching condition. 
The zero eigenfrequency of the topological mode, $E=E_T=0$, in the Eq.(\ref{eq1}) enforces the even eigenfunctions $\psi_{2l}$ ($l=1,2,...$) to be zero irrespective of the existence of the defect coupling $c-g$. Furthermore, for $g=c$ the first equation in Eq. (\ref{eq2}) reduces to
$E_T \psi_{2m}=k\psi_{2m-1}$. Consequently, the only possible choice for $\psi_{2m-1}$ would be zero. Thus, every $\psi_n$ for $n\le2m$ becomes zero meaning that the amplitude of the field at any site up to the defect coupling is zero. This includes the $\psi_0$ meaning that there is no localization at the edge for $g=c$. On the other hand the second equation in the Eq.(\ref{eq2}) for $E=E_T=0$ and the $\psi_{2m}=0$ reduces to similar equation as Eq.(\ref{eq1}a) allowing for a localization at $\psi_{2m+1}$. This explanation nicely matches the observation in the figure (\ref{fig2}). Numerical inspection shows that by increasing the value of $g$ from zero, the $|\psi_1|^2$ starts decreasing and $|\psi_{2m+1}|^2$ starts increasing until they reach to their extremum at $g=c$. This process reverses by making $g>c$ resulting in the recovery of the original topological mode with $E_T=0$ localized at the left edge.

\emph{PT-symmetric defect}--- The anomalous relocation of the topological mode can be observed in a system with only one parity-time (PT)-symmetric defect, using a local gain and loss mechanism as depicted in Fig.(\ref{fig1}c), rather than an asymmetric coupling. Note that while it might seem that the PT-symmetric case is similar to the non-Hermitian coupling the shape of the relocated topological state, in this case, is different from the one generated employing asymmetric coupling. Furthermore, the zero energy mode remains real in the case of asymmetric coupling defect while in the PT-symmetric defect case the zero energy mode becomes imaginary.  

To discuss the relocation of the zero mode systematically, let me start with modifying Eq.(\ref{eq1}) to accommodate for the gain and loss defect. The added gain and loss in the dimer $\psi_{2m}$ and $\psi_{2m+1}$ will change their corresponding equations to
\begin{equation}
\begin{array}{c}
E\psi_{2m}= k \psi_{2m-1} + c \psi_{2m+1}+i\gamma \psi_{2m}\\
E \psi_{2m+1}= c \psi_{2m} + k \psi_{2m+2}-i\gamma\psi_{2m+1}\\
\end{array}
\label{eq3}
\end{equation}
while the rest of the equations in (\ref{eq1}) remain the same. In the above equation, $\mp\gamma$ is the loss ($-$) and gain ($+$) parameter. Such a defect in our system breaks the global PT-symmetry. Consequently for a lattice with a small number of sites a small $\gamma$ can force many modes including the topological mode to become complex. However, as the number of sites increases and (or) defect moves away from the topological edge the energy associated with the topological mode namely $E_T$ stays zero for a larger range of $\gamma$ and becomes imaginary only for a larger value of $\gamma<c$ specifically when the PT-symmetric defect is away from the side where the topological mode is located. 

\begin{figure}
	\centering
	\includegraphics[width=1.\columnwidth]{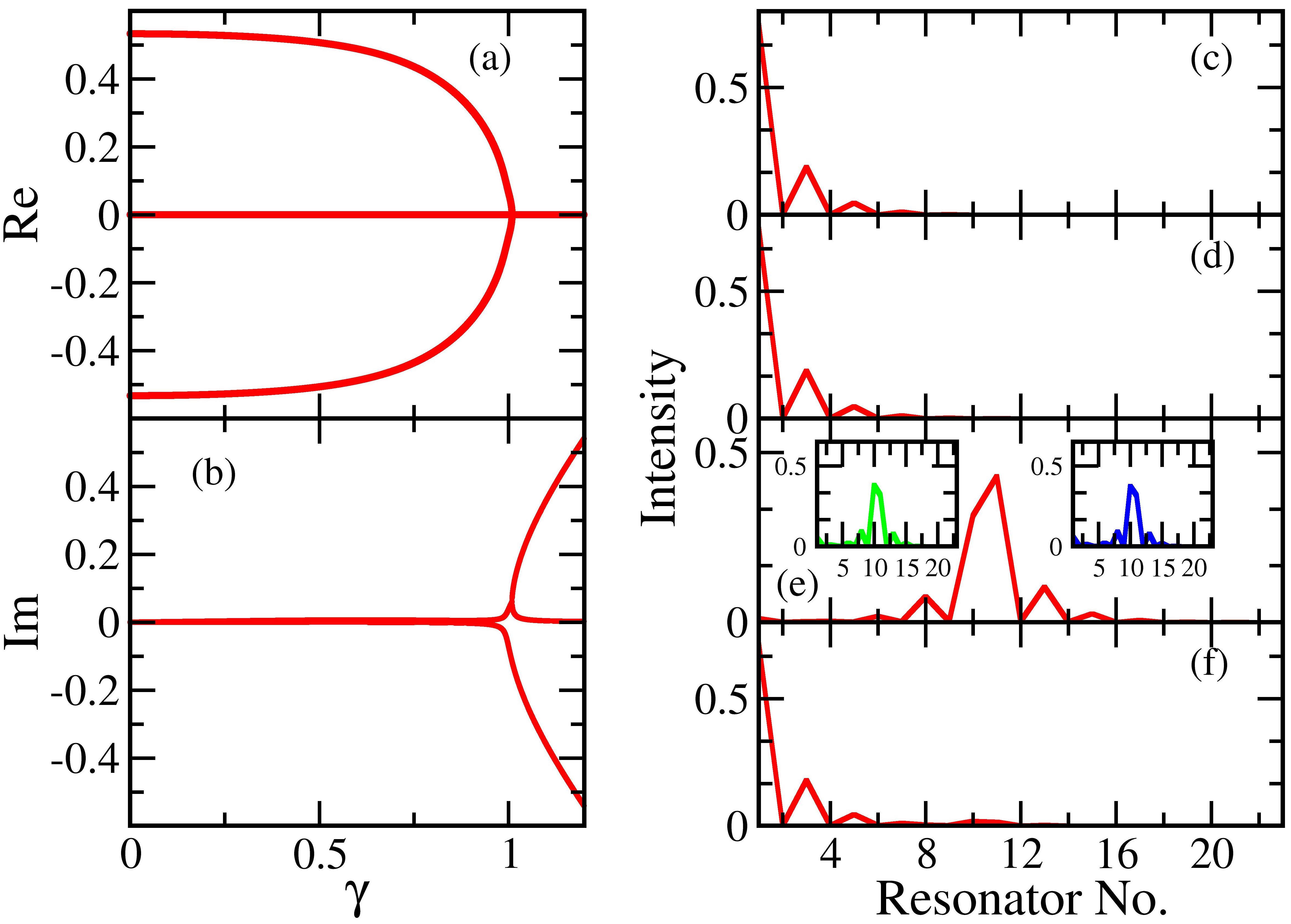}
	\caption{Numerical verification of the topological mode relocation using a PT-symmetric defect depicted in Fig.(\ref{fig1}c). (a) Real part of the energy associated with the topological state with $\text{Re}(E_T)=0$ at the middle and the bound states as a function of $\gamma$ parameter. (b) The same as (a) but the imaginary parts of the energies. We normalized all the parameters to $c=1$ and chose $k=0.5$. At $\gamma\approx 1$ the system has a second-order exceptional point. (c,d) The intensity of the topological state as a function of resonator numbers when $\gamma=0$ and $\gamma=0.2$, respectively. In both cases, the topological state is located at the edge. (e) For $\gamma=1$ the topological state is fully relocated and is localized exponentially around the PT-symmetric defect and with very small intensity at the left edge. The insets show the intensity of the bound states. (f) For $\gamma=1.1>c$ the topological state goes back to the left edge.}
	\label{fig3}
\end{figure}
Figure (\ref{fig3}) summarizes the effect of the PT-symmetric defect in a topological lattice composed of twenty five sites and with couplings $k=0.5$ and $c=1$. The most left coupling is $k$ and thus we expect one topological mode located at the left edge. For the Hermitian case, with $\gamma=0$ we have a topological mode with $E_T=0$ which is localized on the left side, see Fig.(\ref{fig3}a-c). By increasing the strength of the non-Hermiticity the same scenario as the non-Hermitian coupling is observed in Fig.(\ref{fig3}d-f). For  $\gamma$ values smaller than $c$ topological mode stays localized on the left side. As $\gamma$ value approaches $c$ the maximum amplitude of the localized mode intensity relocates to the sites with PT-symmetric defect, as shown in Fig. (\ref{fig3}e), with almost no intensity on the left edge. At this stage, the topological mode is localized around the PT-symmetric defect. Consequently, at this point, there are three localized states around the PT defect state, two associated with the existence of the PT defect and the third one associated with the topological mode (see the inset of Fig. (\ref{fig3}e)). Increasing the gain and loss parameter value beyond $c$ eventually will force the mode to re-appear on the left edge again as depicted in Fig. (\ref{fig3}f).

To explain the relocation of the topological state in a systematic way and for the simplicity of the discussion, let us assume that the PT-symmetric defect is located at the most right side of a lattice with ${\cal N}(\in \text{odd})$ site numbers. In this case Eq.(\ref{eq3}) will become 
\begin{equation}
\begin{array}{cc}
E\psi_{{\cal N}-1}= k \psi_{{\cal N}-2} + c \psi_{{\cal N}}+i\gamma \psi_{{\cal N}-1}&(a)\\
E \psi_{{\cal N}}= c \psi_{{\cal N}-1} -i\gamma\psi_{{\cal N}}&(b)\\
\end{array}
\label{eq4}
\end{equation}
First, we show that unlike the asymmetric coupling scenario at the $\gamma=c$ the $E=E_T$ cannot be zero and is pure imaginary. Let us assume $E_T=0$. From Eq.(\ref{eq4}b) we find that $-i\psi_{{\cal N}-1}=\psi_{{\cal N}}$. By inserting this in Eq.(\ref{eq4}a) we find that $\psi_{{\cal N}-2}$ must be zero. Following the same procedure with other equations in (\ref{eq1}) we find that $\psi_{{\cal N}-2m}$ with $1\leq m\leq \frac{{\cal N}-1}{2}$ must be zero. On the other hand from (\ref{eq1}a) we can show that $\psi_{2m}=0$. Thus, for $E_T=0$ we get the state zero which is a trivial solution meaning that $E_T$ cannot be zero at $c=\gamma$. Now as we expect not to alter the topology of the entire lattice with just one defect, then we would have $\text{Re}(E_T)=0$. Thus, $E_T$ is a purely imaginary number. From Eq.(\ref{eq4}b) we find that $\psi_{{\cal N}-1}=\frac{E_T+ic}{c}\psi_{{\cal N}}$. If we plug this in Eq.(\ref{eq4}a) we find that $\psi_{{\cal N}-2}=\frac{E_T^2}{kc}\psi_{{\cal N}}$. We can follow the same procedure to find the full eigenvector associated with the $E_T$ at $\gamma=c$. Through this process, we find that the eigenstate associated with the $E_T$ would be a decaying function with maximum value at $\psi_{{\cal N}}$. Using similar algebraic calculations we can find other properties associated with the eigenstate associated with the topological state at $\gamma=c$. For example, we can show that $\psi_{2m+1}$ should be real while $\psi_{2m}$ are pure imaginary numbers. This $\pi/2$ phase shift between adjacent sites is a signature of the non-Hermitian topological zero modes.

\begin{figure}
	\includegraphics[width=0.45\columnwidth]{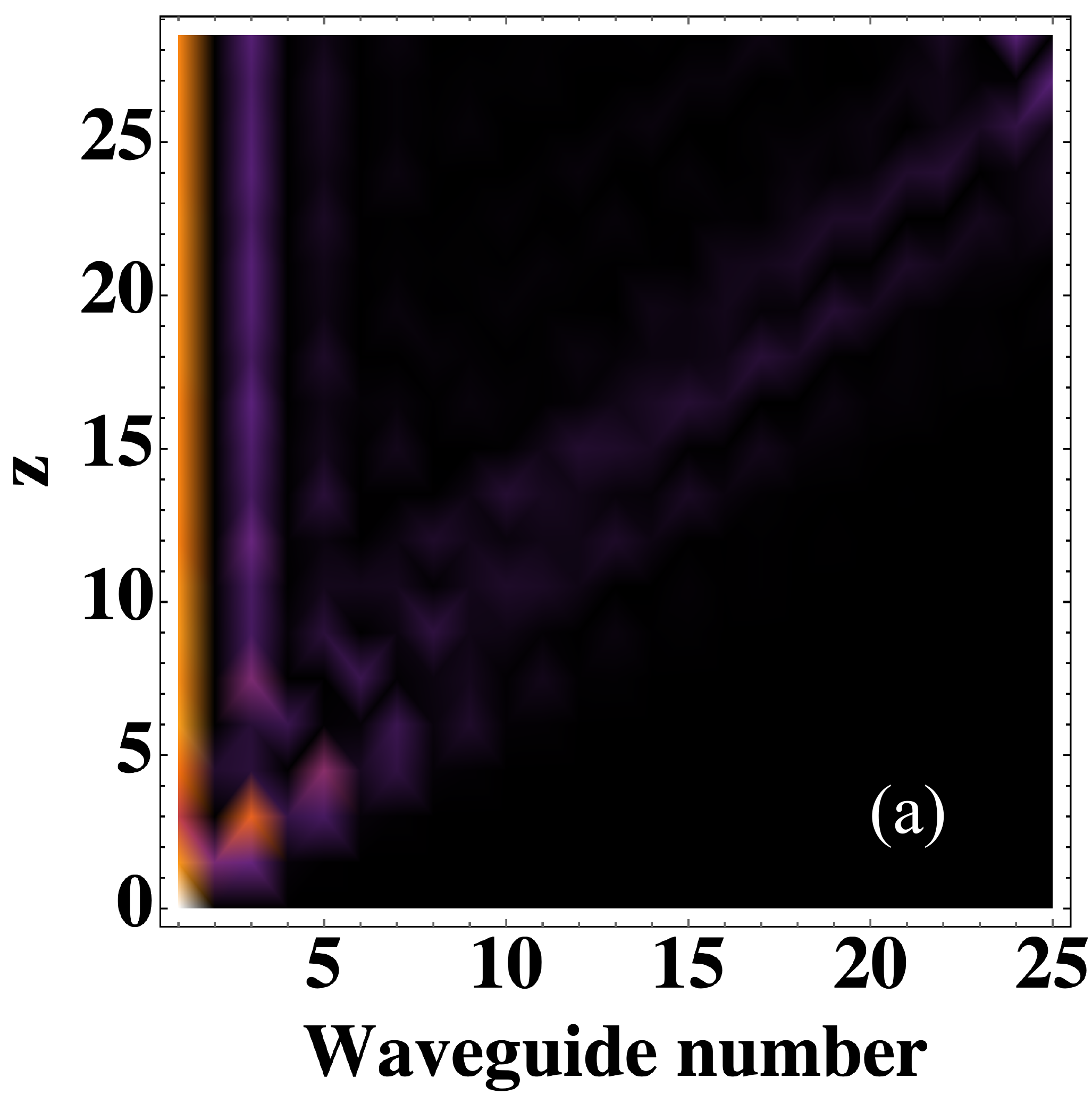}
	\includegraphics[width=0.45\columnwidth]{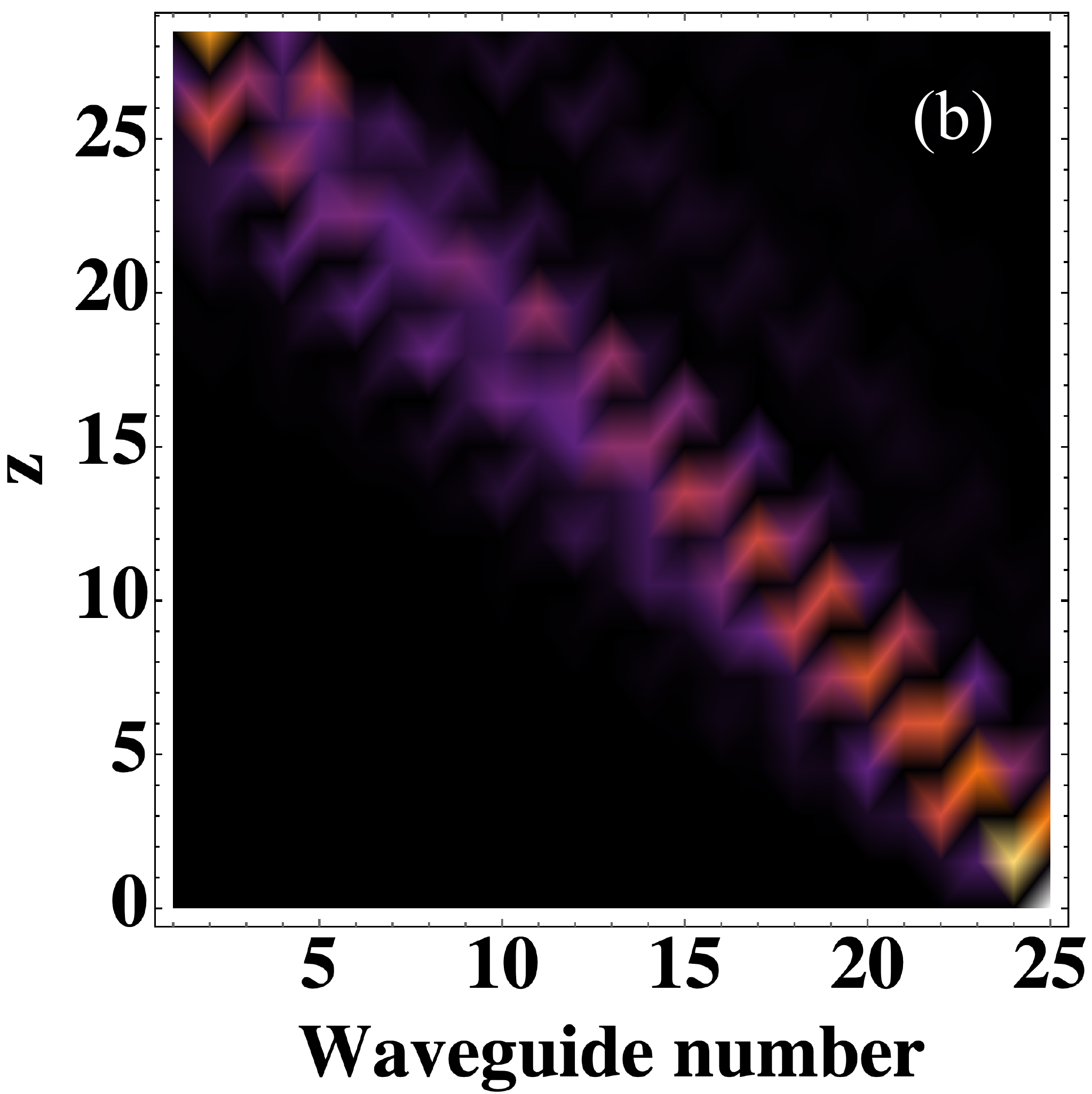}
	\includegraphics[width=0.45\columnwidth]{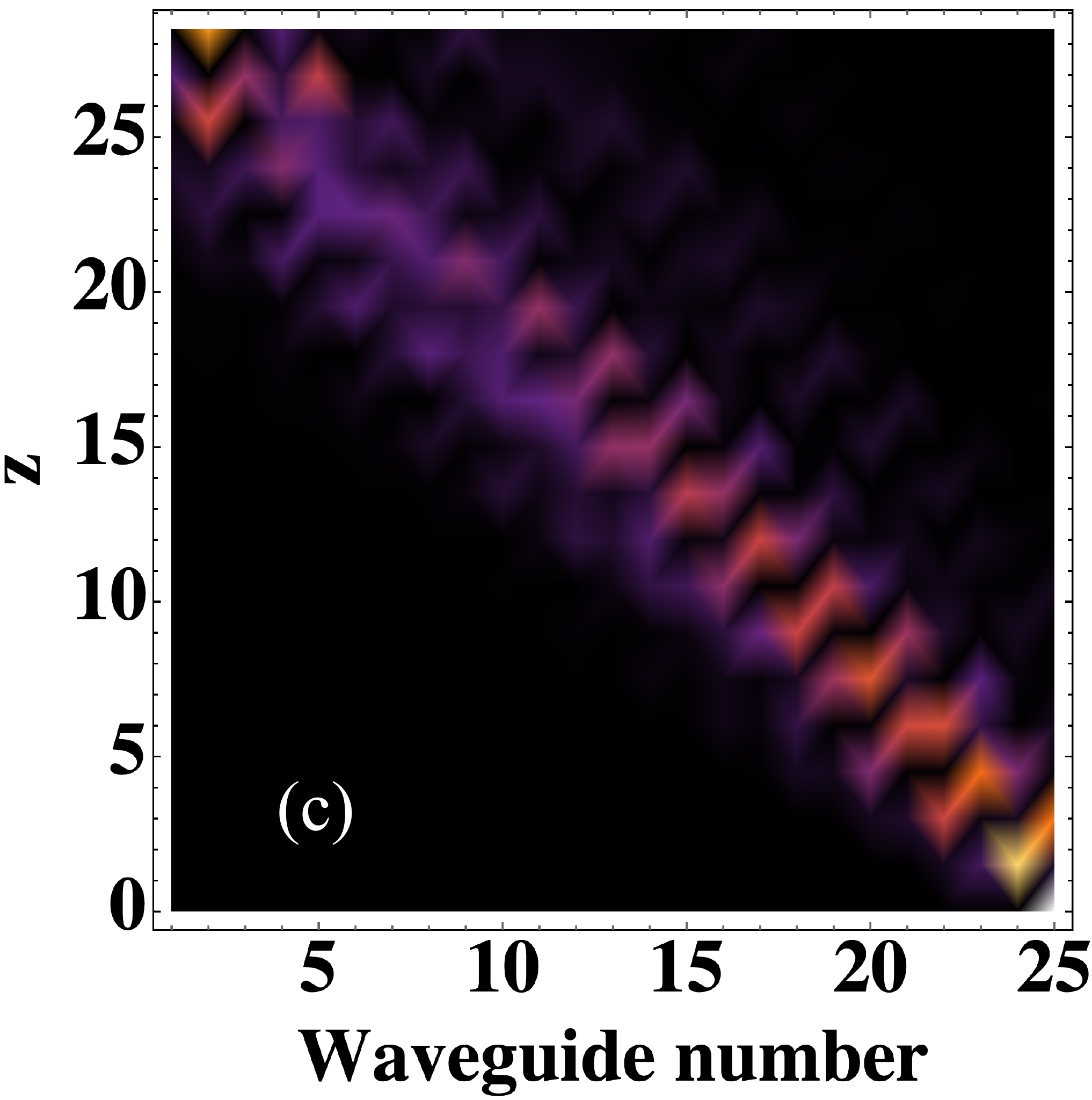}
	\includegraphics[width=0.45\columnwidth]{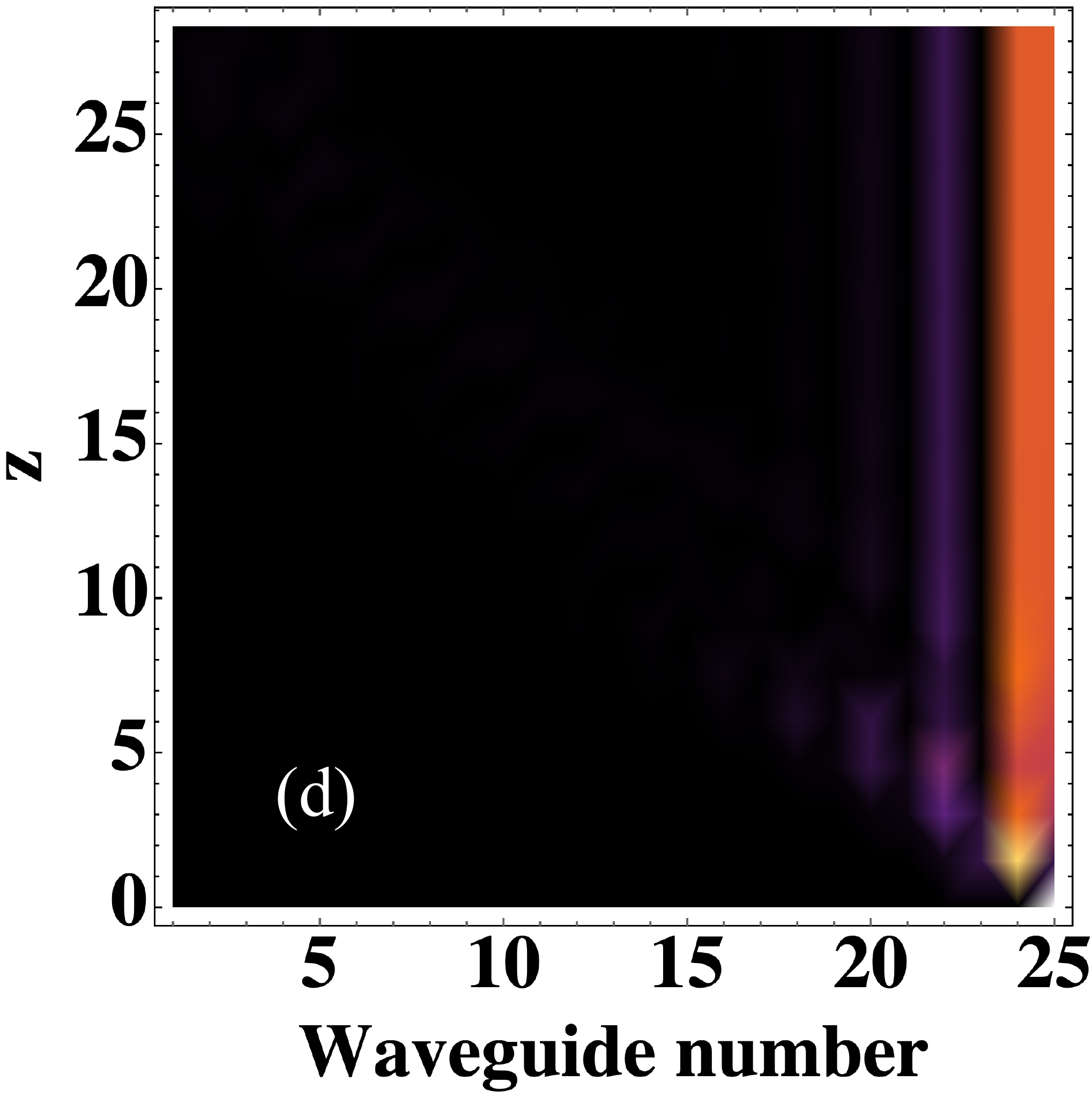}
	\caption{Normalized electric field evolution of a delta function excitation with (a) no defect and excitation at the left edge, (b) no defect and excitation at the right side, (c) gain and loss defect with $\gamma<0.5$ at the right side and excitation at the right side, (d) gain and loss equal to the coupling $c$ and excitation at the right side. At $\gamma=c$ there exists localization at the right edge while in (b,c) no significant localization is observed. }
	\label{fig4}
\end{figure}
We can see the relocation of the topological state using the field propagation in the lattice. In Fig.(\ref{fig4}) we plotted the normalized electric field evolution of a delta function excitation in a lattice made of an array of coupled waveguides arranged to form an SSH lattice. More specifically, we normalized the intensity of the electric field in each waveguide at a specific length $z$ to the total electric field intensity in the lattice at that length $z$ in the unit of larger coupling $c$. Similar to the setup in Fig.(\ref{fig1}a), the waveguides are arranged such that on the left side we expect to have only one topological zero mode while I allow the last two waveguides on the right side to have gain and loss. Figure (\ref{fig4}a) corresponds to the case when there is no gain or loss in the last dimer of the lattice on the right side and we excite the most left waveguide for the sake of comparison. As expected the field remains localized on the left. In Fig. (\ref{fig4}b) I excite the same system from the most right waveguide which shows no localization, which is again expected. In part (c), on the other hand, I increase the gain and loss value to $0<\gamma=0.5c$, and still, no localization is observed for excitation on the most right waveguide. However, when gain and loss in the last two waveguides become equal to the coupling between the two last waveguides, namely $\gamma=c$ the delta function excitation remains localized at the most right waveguide due to the relocation of the topological state.

\emph{Conclusion}--- I have shown that one can remove an edge state from the edge and put it inside the lattice utilizing adding one non-Hermitian defect. While I only consider a 1D structure here, one can extend this idea to 2D and 3D by incorporating line and surface defects similar to what I discussed here, respectively. 


\begin{acknowledgments}
	H.R acknowledges the support by the Army Research Office Grant No. W911NF-20-1-0276 and NSF Grant No. PHY-2012172. The views and conclusions contained in this document are those of the authors and should not be interpreted as representing the official policies, either expressed or implied, of the Army Research Office or the U.S. Government. The U.S. Government is authorized to reproduce and distribute reprints for Government purposes notwithstanding any copyright notation herein. 
	
\end{acknowledgments} 

%

\end{document}